\documentclass[preprint,12pt]{elsarticle}
\journal{Computer Physics Communications}
\usepackage{graphicx}
\usepackage{amsmath}
\usepackage{xspace}
\usepackage{xcolor}
\usepackage{amsmath}
\usepackage[colorlinks=true,urlcolor=blue,linkcolor=black,
  pdfborder={0 0 0}]{hyperref}

\renewcommand{\d}[1][]{\ensuremath{\mathrm{d}^{#1}}\xspace}
\newcommand{\at}[2]{\ensuremath{\left.#1\right|_{#2}}\xspace}
\newcommand{\pt}{\ensuremath{{p_\mathrm{T}}}\xspace}
\newcommand{\sv}{\ensuremath{{\rho}}\xspace}
\newcommand{\urn}{\ensuremath{u}\xspace}
\newcommand{\cil}[1][l]{{\ensuremath{c_{i}^{(#1)}}}\xspace}
\newcommand{\mil}[1][l]{{\ensuremath{m_{i}^{(#1)}}}\xspace}
\newcommand{\sgn}{\ensuremath{\mathrm{sgn}}\xspace}
\newcommand{\cijl}[1][j]{{\ensuremath{c_{i,#1}^{(l)}}}\xspace}
\newcommand{\tev}{\ensuremath{\mathrm{TeV}}\xspace}
\newcommand{\gev}{\ensuremath{\mathrm{GeV}}\xspace}
\newcommand{\kev}{\ensuremath{\mathrm{keV}}\xspace}
\newcommand{\pythia}{\textsc{Pythia}\xspace}
\newcommand{\cimba}{CIMBA\xspace}

\newcommand{\eg}{\textit{e.g.}\xspace}
\newcommand{\ie}{\textit{i.e.}\xspace}
\newcommand{\cmd}[1]{\texttt{#1}\xspace}
\newcommand{\fig}[1]{Figure~\ref{fig:#1}\xspace}
\newcommand{\equ}[1]{Equation~(\ref{equ:#1})\xspace}

\newcommand{\scn}[1]{Section~\ref{sec:#1}\xspace}
\newcommand{\rfr}[1]{Reference~\cite{#1}\xspace}

\newcommand{\keywords}{interpolation, Monte Carlo, event generators, phase
  space and event simulation}

\begin{document}
\begin{frontmatter}
 
\title{\cimba: fast Monte Carlo generation using cubic interpolation}

\author[a]{Philip~Ilten\corref{corauthor}}

\cortext[corauthor]{Corresponding author; 
\textit{e-mail address:} \texttt{philten@cern.ch}}

\address[a]{School of Physics and Astronomy, University of Birmingham,
  Birmingham, UK}

\begin{abstract}
  Monte Carlo generation of high energy particle collisions is a
  critical tool for both theoretical and experimental particle
  physics, connecting perturbative calculations to phenomenological
  models, and theory predictions to full detector simulation. The
  generation of minimum bias events can be particularly
  computationally expensive, where non-perturbative effects play an
  important role and specific processes and fiducial regions can no
  longer be well defined. In particular scenarios, particle guns can
  be used to quickly sample kinematics for single particles produced
  in minimum bias events. \cimba (Cubic Interpolation for Minimum Bias
  Approximation) provides a comprehensive package to smoothly sample
  predefined kinematic grids, from any general purpose Monte Carlo
  generator, for all particles produced in minimum bias events. These
  grids are provided for a number of beam configurations including
  those of the Large Hadron Collider.
\end{abstract}


\end{frontmatter}

\clearpage

\noindent{\bf PROGRAM SUMMARY}\\

\begin{small}
  \setlength{\parindent}{0pt}
  
  \textit{Authors: Philip Ilten}\\

  \textit{Program Title:} \cimba (Cubic Interpolation for Minimum Bias
  Approximation)\\

  \textit{Licensing provisions:} GPL version 2 or later\\

  \textit{Programming language:} Python, C++\\

  \textit{Computer:} commodity PCs, Macs\\

  \textit{Operating systems:} Linux, OS X; should also work on other systems\\

  \textit{RAM:} $\sim$10 megabytes\\


  \textit{Classification:} 11.2 Phase Space and Event Simulation\\

  \textit{Nature of problem:} generation of simulated events in high
  energy particle physics is quickly becoming a bottleneck in analysis
  development for collaborations on the Large Hadron Collider
  (LHC). With the expected long-term continuation of the high
  luminosity LHC, this problem must be solved in the near
  future. Significant progress has been made in designing new ways to
  perform detector simulation, including parametric detector models
  and machine learning techniques, \eg calorimeter shower evolution
  with generative adversarial networks. Consequently, the efficiency
  of generating physics events using general purpose Monte Carlo event
  generators, rather than just detector simulation, needs to be
  improved.\\

  \textit{Solution method:} in many cases, single particle generation
  from pre-sampled phase-space distributions can be used as a fast
  alternative to full event generation. Phase-space distributions
  sampled in particle pseudorapidity and transverse momentum are
  sampled from large, once-off, minimum bias samples generated with
  \textsc{Pythia} 8. A novel smooth sampling of these distributions is
  performed using piecewise cubic Hermite interpolating
  polynomials. Distributions are created for all generated particles,
  as well as particles produced directly from
  hadronisation. Interpolation grid libraries are provided for a number
  of common collider configurations, and code is provided which can
  produce custom interpolation grid libraries.\\

  \textit{Restrictions:} single particle generation\\

  \textit{Unusual features:} none\\

  \textit{Running time:} $\mathcal{O}(10^4)$ particles per second,
  depending on process studied

\end{small}
\clearpage
\sloppy

\section{Introduction}\label{sec:intro}

Monte Carlo simulation of particle physics events provides a critical
tool for calculating theory predictions, understanding detector
performance, designing new detectors, developing experimental analysis
techniques, and understanding experimental
results~\cite{Buckley:2011ms}. Modern general purpose event
generators~\cite{Sjostrand:2014zea, Bothmann:2019yzt, Bellm:2015jjp}
can simulate a broad range of physics processes, from minimum bias
events to specific hard processes such as Higgs production. Final
states from hard processes for a specific fiducial region
corresponding to the acceptance for an experimental analysis can be
selectively and efficiently simulated by choosing the relevant
production channels and limiting the phase-space integration to the
defined fiducial region. For the softer processes which constitute
minimum bias events, such selectivity can no longer be applied at the
perturbative level, as non-perturbative effects begin to have
a much larger impact on the final state particles. Additionally,
simulating minimum bias events can be computationally expensive; an
average Large Hadron Collider (LHC) event can have hundreds of final
state particles.

Consequently, simulating an exclusive final state from minimum bias
events can require significant computing time. For example, $\approx
0.2$ $\phi$-mesons are produced on average per LHC minimum bias event,
with no fiducial requirements. Applying a simple transverse
requirement of $\pt > 1~\gev$ and pseudorapidity requirement of
$|\eta| < 2.5$ reduces this even further to $\approx 0.05$. For more
rare particles such as $B$-mesons or baryons, this suppression can be
orders of magnitude larger, requiring millions of minimum bias events
to be generated to produce a signal candidate. While this is typically
not a problem for the ATLAS and CMS collaborations which focus on
higher \pt physics, this is a significant issue for the LHCb
collaboration, where most of the simulation is produced by extracting
signal candidates from minimum bias simulation~\cite{Belyaev:2011zza}.

\begin{figure}
  \includegraphics[width=0.5\textwidth]{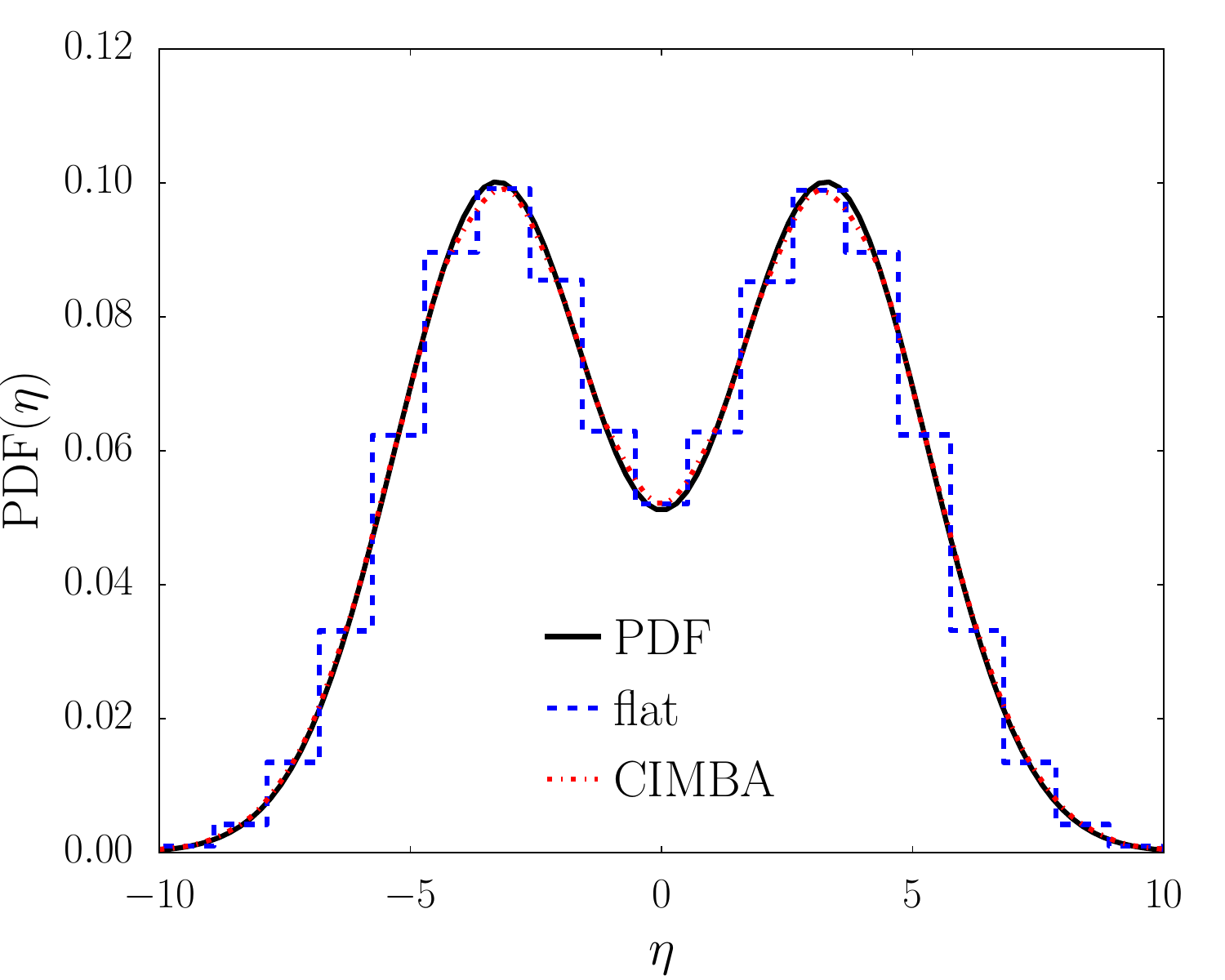}
  \includegraphics[width=0.5\textwidth]{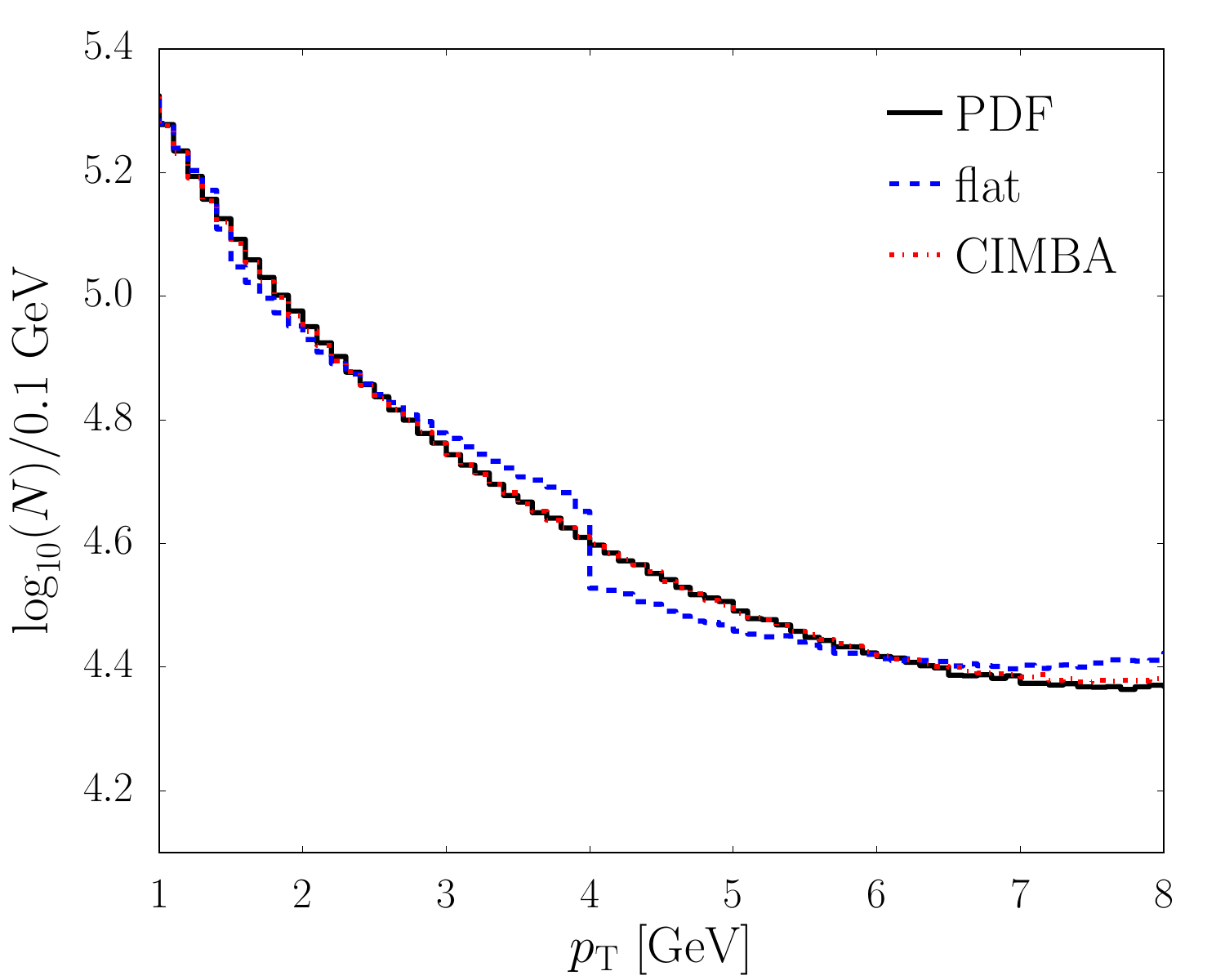}
  \caption{Simplified example of (left) a generating PDF and (right) a
    distribution generated from sampling the
    PDF.\label{fig:artifacts}}
\end{figure}

When only the signal candidate is needed, a particle gun technique can
be used to more efficiently generate signal, where a probability
density function (PDF) is used to quickly sample phase-space for the
signal particle. Typically, this PDF is defined as a function of one
or two variables, and is produced from a histogram, although other
non-parametric estimation and simulation techniques could be used
instead, see \eg \rfr{Thompson:1990zz} for an overview. The sharp
edges of the histogram can produce artefacts in generated events. In
\fig{artifacts} a semi-realistic example is given, where the PDF for
an arbitrary signal particle is given as a function of
pseudorapidity. The actual closed form of this PDF is rarely known,
but here a simplified double Gaussian distribution has been
used. Assuming a fixed momentum of $10~\gev$, the generated transverse
momentum distribution is given, sampled from the actual PDF and the
histogram PDF, resulting in a sharp edge at a transverse momentum of
$\approx 4~\gev$.

The \cimba package (Cubic Interpolation for Minimum Bias
Approximation), available at
\href{https://gitlab.com/philten/cimba/}{\cmd{gitlab.com/philten/cimba/}},
provides a complete framework for generating particles produced in
minimum bias events, using a careful choice of generating PDF
variables and smooth interpolation of the PDF. In the example of
\fig{artifacts}, this smooth interpolation of the histogram PDF
results in a nearly identical distribution to the true underlying
distribution. The methods developed in this paper for sampling both
univariate and bivariate interpolated distributions can also be
applied outside the context of particle physics, where smooth sampling
of discrete distributions is required. The remainder of this paper is
organised as follows. In \scn{ui} the choice of univariate
interpolation used for \cimba is explained. In \scn{us} the method for
sampling this interpolation is derived and in \scn{bs} a method for
sampling an interpolated bivariate PDF is introduced. The signal
generation method for \cimba is given in \scn{pgun}, while the
structure of the \cimba package is outlined in \scn{sa}. Some
examples, along with validation and performance, are explored in
\scn{vpa}, and the paper is concluded in \scn{con}.

\section{Univariate Interpolation}\label{sec:ui}

The piecewise interpolating polynomials, $f(x)$, for a PDF defined by
a histogram, $h(x)$, with $n$ bin centres $x_i$ and bin values $y_i$
is given by,
\begin{equation}
  f_i(x) = \sum_{l = 0}^{k_i} \cil (x - x_i)^l,
  \label{equ:pdf}
\end{equation}
given $x_{i} \leq x < x_{i + 1}$, and $k_i$ is the degree of the
interpolating polynomial $f_i$. For $f(x)$ to be both continuous and
smooth, \eg $C^1$, then $k \geq 3$. For most applications, $k = 3$ is
sufficient; in many cases using $k > 3$ produces worse results due to
Runge's phenomenon~\cite{Runge:1901zz}.

For a histogram with $n \geq 3$, \eg at least three bins, cubic
interpolation can be performed with $n - 1$ cubic functions $f_i$,
resulting in $4(n - 1)$ unknown coefficients $\cil$. A number of
requirements, dependent on the interpolation method, must be placed on
$f_i$ to determine $\cil$. In most methods, the piecewise
interpolating function is required to be continuous,
\begin{equation}
    f_i(x_i) = h(x_i) ~\textrm{and}~ f_{i}(x_{i + 1}) = h(x_{i + 1}),
\end{equation}
which specifies $2(n - 1)$ constraints.

Cubic Hermite interpolation requires the first derivative at each
point must also match,
\begin{equation}
  \at{\frac{\d{f_i}}{\d{x}}}{x_i} = \at{\frac{\d{h}}{\d{x}}}{x_i}
  ~\mathrm{and}~
  \at{\frac{\d{f_i}}{\d{x}}}{x_{i + 1}} = \at{\frac{\d{h}}{\d{x}}}{x_{i + 1}},
\end{equation}
providing an additional $2(n - 1)$ constraints. However, this requires
the first derivative is known at each $x_i$. Monotone cubic Hermite
interpolation (PCHIP) is the same as cubic Hermite interpolation,
except the first derivative equality is relaxed when necessary to
ensure the interpolating polynomial is monotonic over its interval of
validity~\cite{Fritsch:1980zz}.

The unknown first derivative of $h(x)$ may be approximated as the
average secant of the two neighbouring intervals for a given $x_i$,
\begin{equation}
  \at{\frac{\d{h}}{\d{x}}}{x_i} \approx \frac{1}{2} \left(
  \frac{y_i - y_{i - 1}}{x_i - x_{i - 1}} +
  \frac{y_{i + 1} - y_i}{x_{i + 1} - x_i}
  \right),
\end{equation}
except for the endpoints $x_1$ and $x_n$. The derivative for these
endpoints can be approximated using extrapolation. Assuming linear
extrapolation, the first derivative is then defined as,
\begin{equation}
  \at{\frac{\d{h}}{\d{x}}}{x_1} \approx \frac{y_2 - y_1}{x_2 - x_1}
  ~\mathrm{and}~
  \at{\frac{\d{h}}{\d{x}}}{x_n} \approx \frac{y_n - y_{n -1}}{x_n - x_{n -1}},
\end{equation}
for $x_1$ and $x_n$.

Natural cubic interpolation does not require knowledge of the first
derivative of $h(x)$, but instead requires the first and second
derivatives of the interpolating function are smooth,
\begin{equation}
  \at{\frac{\d{f_i}}{\d{x}}}{x_i} =
  \at{\frac{\d{f_i}}{\d{x}}}{x_{i + 1}}
  ~\mathrm{and}
  \at{\frac{\d[2]{f_i}}{\d{x}^2}}{x_i} =
  \at{\frac{\d[2]{f_i}}{\d{x}^2}}{x_{i + 1}},
\end{equation}
\eg $f(x)$ is $C^2$. This provides an additional $2(n - 2)$
constraints. The final two constraints,
\begin{equation}
  \at{\frac{\d[2]{f_1}}{\d{x}^2}}{x_1} = 0
  ~\mathrm{and}~
  \at{\frac{\d[2]{f_{n - 1}}}{\d{x}^2}}{x_n} = 0,
\end{equation}
are required to fully specify the coefficients for $f_i$. Alternative
constraints are the clamped boundary condition, where the first
derivatives are required to be zero, and the not-a-knot boundary
condition where the $x_2$ and $x_{n-1}$ third derivatives are required
to be continuous.

\begin{figure}
  \includegraphics[width=0.5\textwidth]{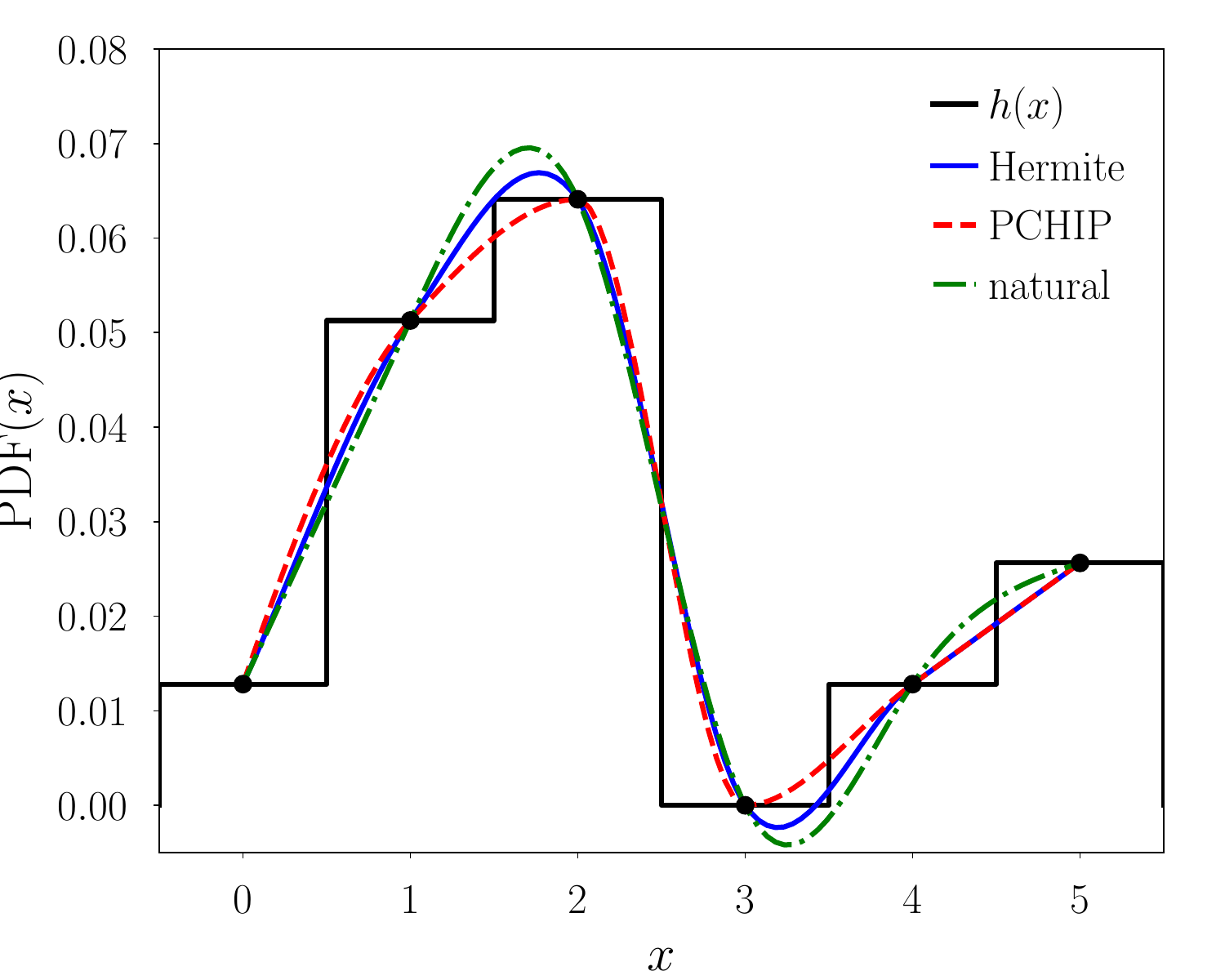}
  \includegraphics[width=0.5\textwidth]{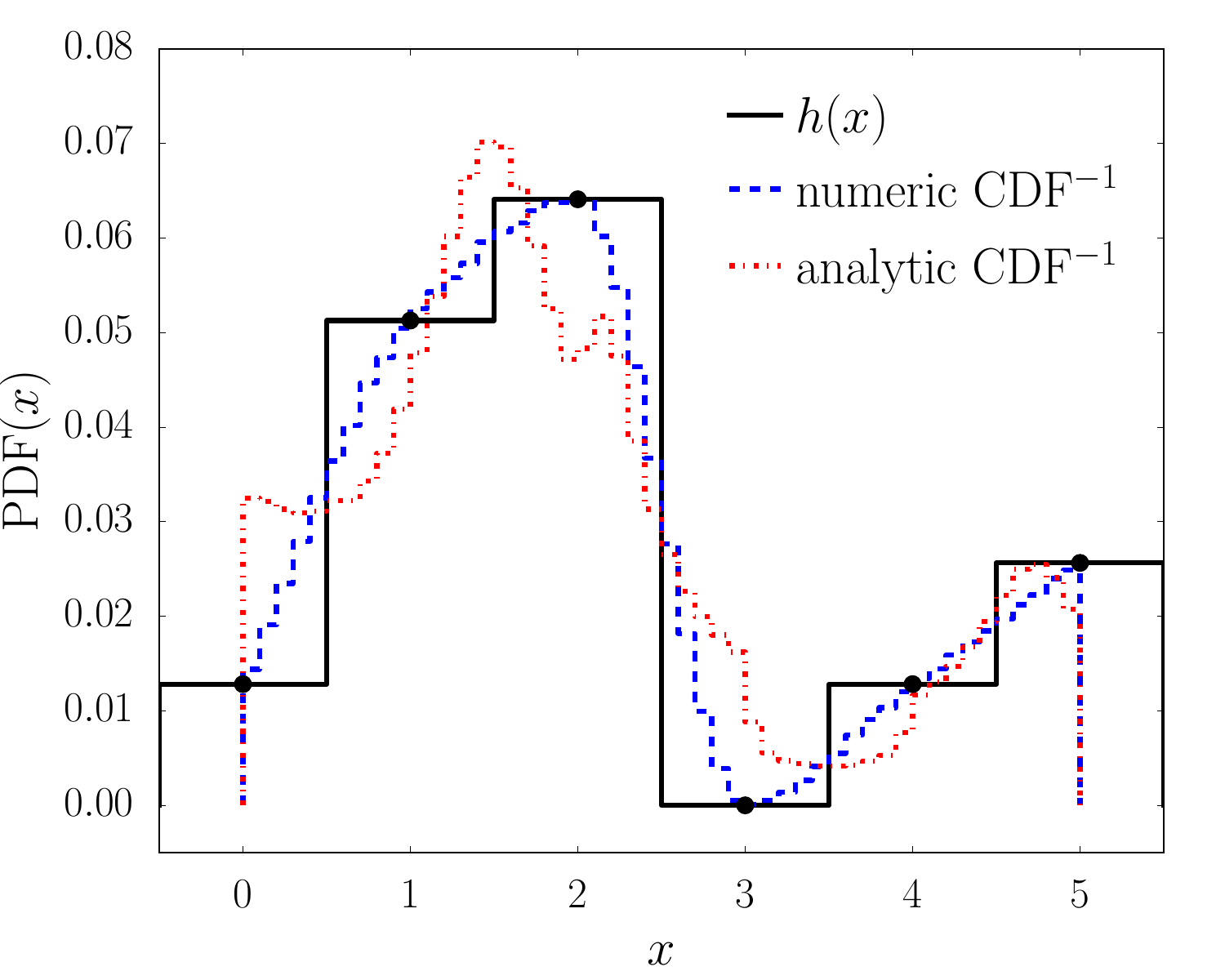}
  \caption{(left) Different methods of cubic interpolation and (right)
    demonstration of poor behaviour in the sampled PDF when
    interpolating the inverse CDF.\label{fig:methods}}
\end{figure}

An example of Hermite, PCHIP, and natural cubic interpolation of a
histogram $h(x)$ defining a PDF is given in \fig{methods}. While the
normalisation of $h(x)$ is unity, this is not the case for the
interpolating functions. Both the natural and Hermite interpolations
fluctuate around the histogram bin centres, while the PCHIP
interpolation does not. More importantly, because the PCHIP is
monotonic and must pass through each bin centre, the resulting
interpolation is guaranteed to always be positive. Neither the natural
nor Hermite methods can guarantee a positive interpolation, making
these two methods unsuitable for creating interpolated PDFs of
histograms, and so the PCHIP method is used instead. Ideally, a
modified PCHIP could be used, where the integral of the interpolating
polynomial over each histogram bin yields the bin content.

\section{Univariate Sampling}\label{sec:us}

The PDF $g(x)$, defined over the interval $x_{\min}$ to $x_{\max}$, can be
randomly sampled from a uniform distribution, $\urn \in [0, 1]$, by
taking $G^{-1}(u)$, where
\begin{equation}
  G(x) = \int_{x_{\min}}^x \d{x'}\,g(x'),
\end{equation}
is the cumulative distribution function (CDF) of $g(x)$ and
$G^{-1}(u)$ is the inverse of this CDF. Naively, rather than
analytically inverting $G(x)$ one could sample $G(x)$ at $x_i$,
numerically invert the function by swapping $G(x_i)$ with $x_i$, and
interpolate the resulting points. However, if cubic interpolation of
$G^{-1}(u)$ is used, the generated PDF is defined by second-order
polynomials and typically has very poor behaviour, as shown in
\fig{methods}. Consequently, analytic inversion of $G(x)$ is
necessary.

The piecewise CDF of the interpolating functions $f_i(x)$ from
\equ{pdf} is defined iteratively,
\begin{equation}
  F_i(x) = F_{i - 1}(x_i) + \sum_{l = 0}^{k_i} \frac{\cil}{l + 1}
  (x - x_i)^{l + 1},
  \label{equ:cdf}
\end{equation}
where $\cil$ are from the already normalised $f_i(x)$ and $F_0 =
0$. In practice, it is easier to use the coefficients from the
unormalised $f(x)$ and sample $\urn \in [0, F_{n - 1}(x_n)]$ rather
than over the unit interval.

To randomly sample this distribution $F_{i}^{-1}(u)$, the inverse of $F_i(x)$ must be
taken, which requires solving up to a fourth-order polynomial,
\begin{equation}
  0 = F_{i-1}(x_i) - u + \sum_{l = 0}^{k_i} \frac{\cil}{l +
    1} (x - x_i)^{l + 1}
\end{equation}
given $k = 3$ and $F_i(x) = u$. For simplicity, the monic form of this
polynomial is solved,
\begin{equation}
  0 = (x - x_i)^{k_i} + \sum_{l = 0}^{k_i}\mil(x - x_i)^l,
  \label{equ:monic}
\end{equation}
where $k_i$ is the highest non-zero order, $\mil[0] = F_{i-1}(x_i) -
4u/\cil[3]$, and $\mil = 4\cil[l - 1]/(l\cil[k])$ for $1 \leq l \leq
k$.

In the linear case there is a single solution to \equ{monic}
\begin{equation}
  F_{i}^{-1}(u) = -\mil[0]/\mil[1],
  \label{equ:linear}
\end{equation}
while for higher orders with multiple roots, the correct solution is the positive root closest to zero. For the quadratic case there are always two real roots and so,
\begin{equation}
  F_{i}^{-1}(u) =
  \begin{cases}
    q &
    \text{if $q \geq 0$ and $q^2 > \mil[0]$} \\
    \mil[0]/q &
    \text{else,} \\
  \end{cases}
  \label{equ:quadratic}
\end{equation}
where $q = -(\mil[1] + \sgn(\mil[1])\sqrt{\mil[1]^2 - 4\mil[0]^2})/2$. For the cubic case there can be either a single or three real roots,
\begin{equation}
  F_{i}^{-1}(u) =
  \begin{cases}
    s + q/s - \mil[2]/3 &
    \text{if $p^2 > q^3$} \\
    2\sqrt{q}\cos((\cos^{-1}(p/\sqrt{q^3}) - 2\pi)/3) - \mil[2]/3 &
    \text{else,} \\
  \end{cases}
  \label{equ:cubic}
\end{equation}
where $s = \sgn{(p)}(|p| + \sqrt{p^2 - q^3})^{1/3}$, $p = (2\mil[2]^3 -
  9\mil[2]\mil[1] + 27\mil[0])/54$, and $q = (\mil[2]^2 -
      3\mil[1])/9$. Finally, for the quartic case,
\begin{equation}
  F_{i}^{-1}(u) = -t(s - \sqrt{tq/s - p - 4s^2}/2) - \mil[3]/4,
  \label{equ:quartic}
\end{equation}
given $p = (8\mil[2] - 3\mil[3]^2)/4$ and $q = (\mil[3]^3 -
    4\mil[3]\mil[2] + 8\mil[1])/8$. When $d_1^2 > 4d_0^3$,
\begin{equation}
  \begin{aligned}
    d_0 &= \mil[2]^2 - 3\mil[3]\mil[1] + 12\mil[0] \\
    d_1 &= 2\mil[2]^3 - 9\mil[3]\mil[2]\mil[1] + 27\mil[3]^2\mil[0] +
           27\mil[1]^2 - 72\mil[2]\mil[0] \\
  \end{aligned}
\end{equation}
there are two real roots with the correct solution given by,
\begin{equation}
  \begin{aligned}
    r &= ((d_1 + \sqrt{d_1^2 - 4d_0^3})/2)^{1/3} \\
    s &= \sqrt{(r + d_0/r - p)/3}/2 \\
    t &= \begin{cases}
      -1 & \text{if $q/s - p - 4s^2 < 0$} \\
      1 & \text{else.} \\
    \end{cases} \\
  \end{aligned}
\end{equation}
Otherwise, there are four real roots, with the correct solution given
by,
\begin{equation}
  \begin{aligned}
    s &= ((\sqrt{d_0}\cos(\cos^{-1}(d_1/(2d_0^{3/2}))/3) - p/2)/6)^{1/2} \\
    t &= \begin{cases}
      -1 & \text{if $\mil[3]/4 + s > \sqrt{q/s - p - 4s^2}/2$} \\
      1 & \text{else.} \\
    \end{cases} \\
  \end{aligned}
\end{equation}
The quadratic and cubic solutions are well conditioned
numerically~\cite{Press:2002zz}, while the quartic solution is
not~\cite{HerbisonEvans:1995zz}. Currently, there is no numerically
stable closed form for the quartic solution, although a generalised
form of the cubic solvent is available, and could potentially be used
to produce a numerically stable solution~\cite{Shmakov:2011zz}. When
the solution of \equ{quartic} fails numerically, a result will
oftentimes be returned outside the interpolation interval and a more
stable, albeit slower method, \eg finding the eigenvalues of the
companion matrix~\cite{Horn:2012zz}, can be used instead.

\section{Bivariate Sampling}\label{sec:bs}

A bivariate PDF $g(x, y)$ can be sampled by factorising the generation
into two univariate samplings. First project the PDF onto a single
variable,
\begin{equation}
  g(x) = \int_{y_{\min}}^{y_{\max}} \d{y}\, g(x, y)
\end{equation}
and build the CDF for this projection, $G(x)$. A value for $x$ can
then be generated using univariate sampling, \eg from
$G^{-1}(u)$. Given this $x$, a value for $y$ can be sampled with
$G^{-1}(v|x)$ from a uniform distribution $v \in [0, 1]$, where
\begin{equation}
  G(y|x)= \int_{y_{\min}}^{y} \d{y'}\, g(x, y')
\end{equation}
and $G^{-1}(v|x)$ is the inverse of $G(y|x)$. Again, this can be
selected using univariate sampling.

Consider a bivariate PDF defined by a histogram $h(x, y)$ with $n$
$x$-bin centres of $x_{i,j}$, $m$ $y$-bin centres of $y_{i,j}$, and $n
\times m$ bin contents $z_{i,j}$. Given PCHIP interpolation $f(x,y)$
of $h(x,y)$, the projected PDF $f(x)$ can be factored into arbitrary
non-polynomial terms $\zeta_{i,j}$,
\begin{equation}
  \sum_{j = 1}^{m - 2}\left(y_{i,j - 1} - y_{i,j + 1}\right) \left[(y_{i,j +
      1} - y_{i,j}) - (y_{i,j + 1} - y_{i,j - 1})\right]\zeta_{i,j}(x),
\end{equation}
and polynomial terms for each $f_i(x)$. The term in square brackets is
the difference between two bin centres in $y$, so if the histogram has
constant bin-spacing $\Delta y$, the $\zeta_{i,j}$ terms vanish,
resulting in a third-degree polynomial to which the method of \scn{us}
can be applied.

Assuming constant $\Delta y$, the projected PDF is given by,
\begin{equation}
  f_i(x) = \frac{\Delta y}{12}\sum_{l = 0}^{k} x^l
  \begin{cases}
    5\cijl[0] + 14\cijl[1] + 5\cijl[2]
    & \text{if $m = 3$,} \\
    5\cijl[0] + 13\cijl[1] + 13\cijl[2] + 5\cijl[3]
    & \text{if $m = 4$,} \\
    5\cijl[0] + 13\cijl[1] + 13\cijl[m - 2] + 5\cijl[m - 1] +
    \sum_{j = 2}^{m - 3} 12 \cijl
    & \text{else,} \\
  \end{cases}
  \label{equ:projection}
\end{equation}
where \cijl are the coefficients for the interpolating function
$f_i(x|j)$ for a given $y$-bin $j$. For example, $f_i(x|0)$ are the
interpolating functions defined by the points $x_{i,0}$ and $z_{i,0}$
for the first $y$-bin with $j = 0$. Consequently, coefficients for the
$m$ interpolating functions $f_i(x|j)$ must be built to determine the
coefficients for $f_i(x)$.

Using \equ{projection} and the method of \scn{us}, an $x$-value can
now be sampled. Given this $x$, the $f_i(x|j)$ can be used to
calculate interpolated $z_j$ for each $y_j$. From these points the
cubic interpolation function $f_j(y|x)$ can be sampled using the same
process as for $x$. Note that to sample $y$ after an $x$ is given,
$4m$ interpolation coefficients must be determined from $m$
interpolated values. This is computationally more expensive than just
performing bicubic interpolation, which only requires $16$
interpolation coefficients from $4$ interpolated values.

\section{Single Particle Generation}\label{sec:pgun}

For the specific application of \cimba, \ie single particle generation
from minimum bias events, each particle can be fully specified with a
three-momentum, $(p_x, p_y, p_z)$, and a mass, $m$. The mass of the
particle is independent of its momentum, and can be sampled from a
line-shape distribution given the nominal mass and width of the
particle, leaving three unknowns. For colliders with unpolarised
beams, particles are produced uniformly in azimuthal angle $\phi$ in
the centre-of-mass (COM) frame, and so $\phi$ can also be sampled
independently. This leaves two dependent variables which can be
generated using the bivariate sampling of \scn{bs}.

To minimise the effects of histogram binning, these two dependent
variables should be chosen to have as flat of distributions as
possible. In the COM frame, particle production is relatively constant
in pseudorapidity, unlike the polar angle $\theta$, making this a
natural choice. The remaining variable can then be either the
magnitude of the momentum, $p$, or the transverse momentum, \pt. At
the LHC, particle production occurs primarily through $t$-channel
processes, \eg $gg \to q\bar{q}$ and $qq' \to qq'$, which are
proportional to $\pt^{-4}$. However, these processes are convolved
with the proton PDF, modifying the \pt distribution.

\begin{figure}
  \includegraphics[width=0.5\textwidth]{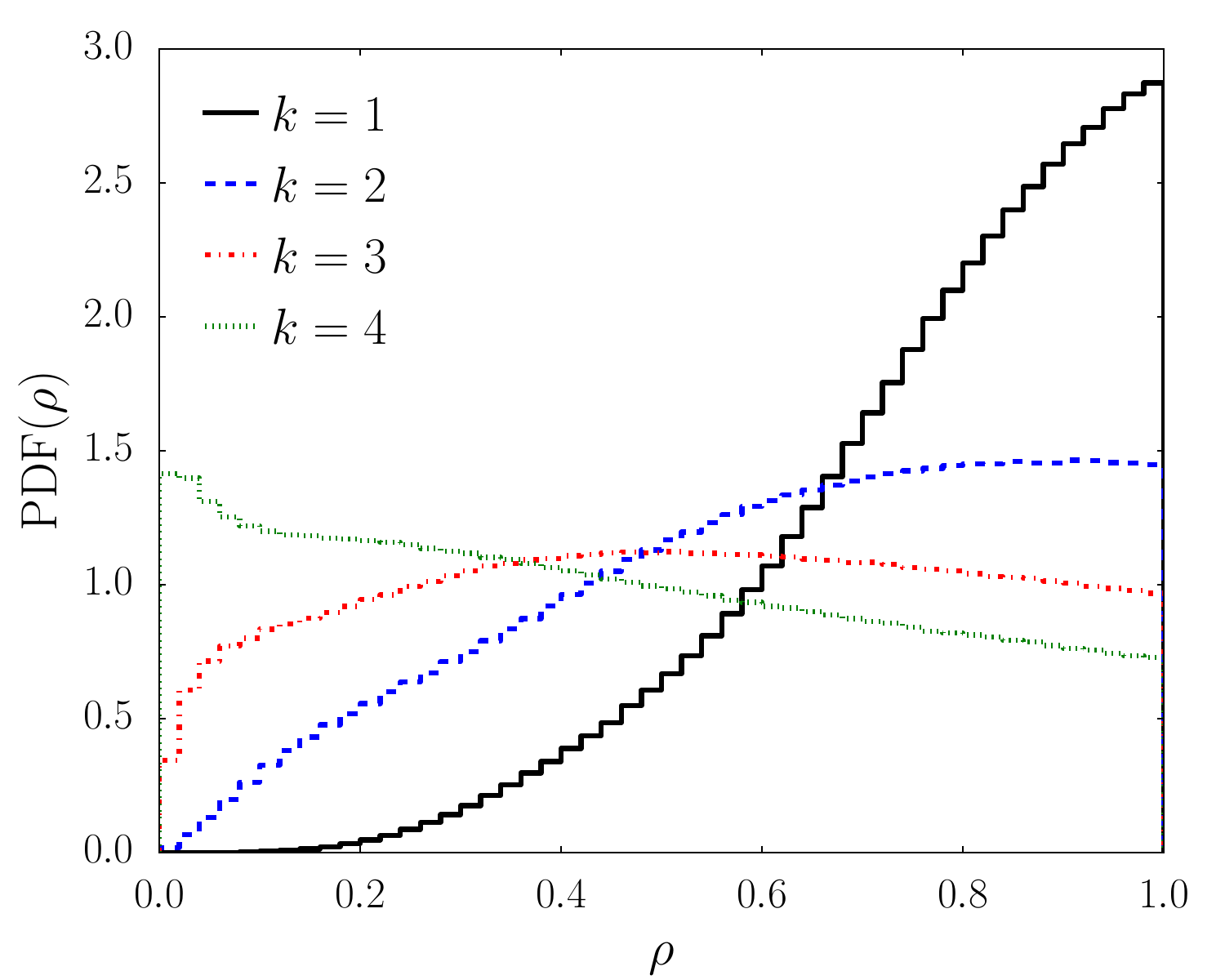}
  \includegraphics[width=0.5\textwidth]{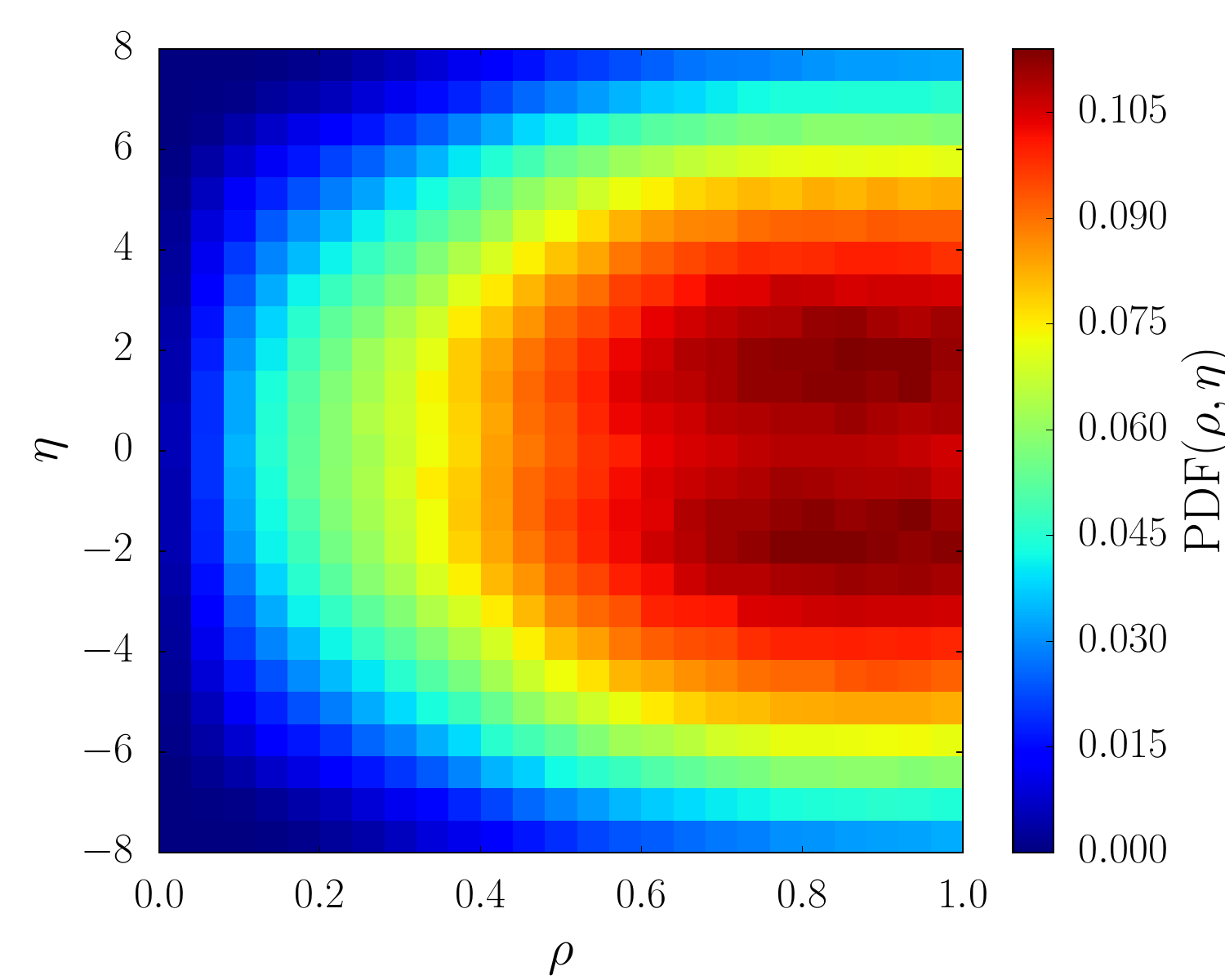}
  \caption{(left) PDF of $\rho$ for various $k$ and (right) bivariate
    PDF in $\rho$ ($\pt_{\min} = 0.25~\gev$ and $k = 2$) and
    pseudorapidity, given for all final particles in LHC minimum bias
    events.\label{fig:grids}}
\end{figure}

The second variable is chosen as,
\begin{equation}
  \sv \equiv \frac{1}{(\pt + 1~\gev - \pt_{\min})^k}
\end{equation}
where $\pt_{\min}$ is the minimum \pt requirement on produced
particles, and $k$ is an arbitrary power. This variable is relatively
flat, bounded between $0$ and $1$, the PDF must be $0$ at $\rho = 0$,
and can be reasonably calculated at $\rho = 1$ via linear
extrapolation. By default in \cimba, $\pt_{\min} = 0.25~\gev$ and $k =
2$, although new grids can be generated by the user with different
values. Here, $k = 2$ is chosen as it provides a PDF that can smoothly
interpolated to $0$ for $\rho = 0$ and linearly extrapolated to $\rho
= 1$. The $\pt_{\min}$ value is chosen for LHCb applications, where
values below this cut-off are not experimentally accessible. In
\fig{grids} the projected PDF for $\rho$ with various $k$ and the
bivariate PDF for $\rho$ and pseudorapidity is given for all final
particles in LHC minimum bias events.

The default interpolation grids in pseudorapidity and $\rho$ are
generated for the LHC using \pythia 8.240~\cite{Sjostrand:2014zea}
with the \cmd{SoftQCD:all = on} flag. Code is provided in \cimba so
that additional grids can be generated with \pythia for any COM energy
and beam setup, \eg proton-lead or proton-electron, and an arbitrary
user configuration of \pythia. Regularisation of low \pt divergences
for processes such as $gg \to b\bar{b}$ is performed using the same
damping \pythia applies to soft QCD events. For some beam setups, like
lepton-lepton, the selected physics processes need to modified as soft
QCD production is no longer relevant. Additionally, user provided
grids from other event generators can also be passed to \cimba,
assuming they are in the same format as the \pythia grids.

For each configuration, two sets of grids are provided.
\begin{itemize}
\item All particles produced in the event, including intermediate
  states. Only the final version of duplicate particles are included,
  \eg for a gluon which undergoes multiple splittings, only the gluon
  from the final splitting is kept.
\item Only particles produced directly from the hadronisation.
\end{itemize}
Because \pythia performs $B$-oscillations during the decay stage of
the event generation, these are switched off to ensure double
oscillations do not occur when passing generated $B$-mesons back to
\pythia for subsequent decays.

\section{Program Structure and Algorithms}\label{sec:sa}

\begin{figure}
  \includegraphics[width=\textwidth]{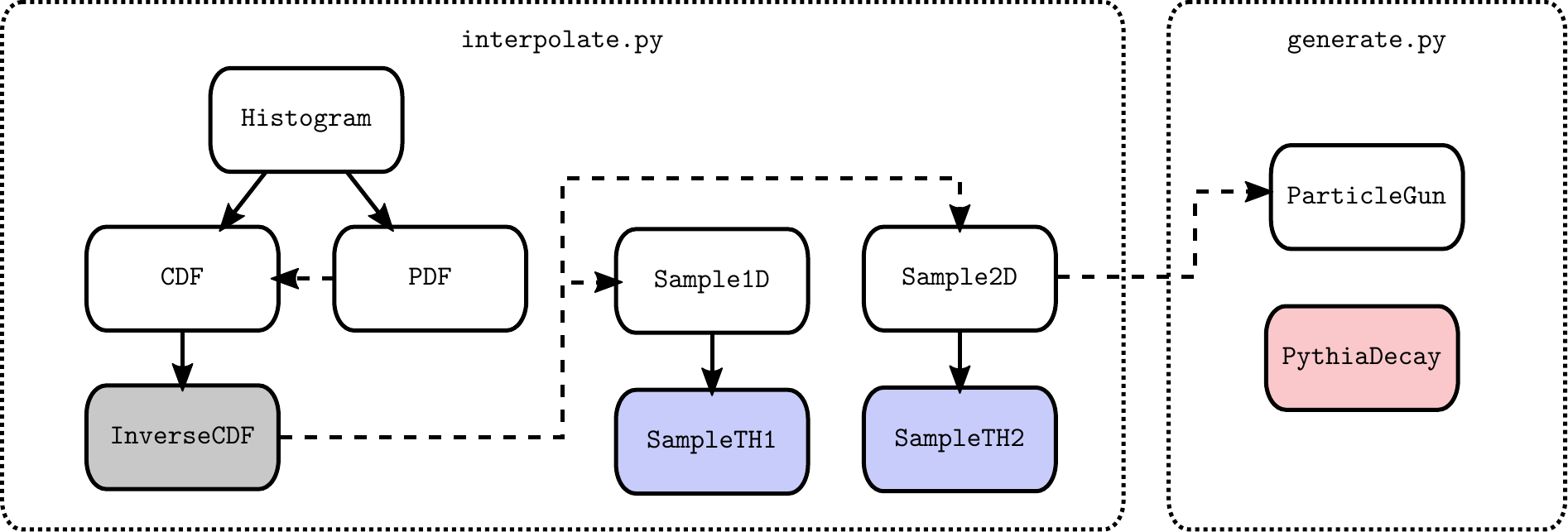}
  \caption{Structure of the \cimba package with the dotted border
    boxes representing files and the solid border boxes representing
    classes. Solid arrows indicate inheritance, while dashed arrows
    indicate dependence. Colours indicate external module
    dependencies: grey is an optional \cmd{numpy} dependence, blue is
    a \cmd{ROOT} dependence, and red is a \cmd{pythia8}
    dependence.\label{fig:structure}}
\end{figure}

The structure of the \cimba package is summarised in \fig{structure}
where the class \cmd{Histogram} provides fast one dimensional bin
finding using a modified regula falsi method; if regula falsi fails to
find a bracketing interval, then the bisection method is used
instead. The \cmd{Histogram} class acts as the base class for the
\cmd{PDF}, \cmd{CDF}, and \cmd{InverseCDF} interpolation classes. The
\cmd{PDF} class performs cubic interpolation using the PCHIP method
outlined in \scn{ui}, given an initial set of interpolating points. If
provided with a set of \cmd{PDF} objects and corresponding $y$-values,
the coefficients for the projected PDF of \equ{projection} are built
and used instead. The \cmd{CDF} class is initialised with a \cmd{PDF}
object and returns the CDF defined by \equ{cdf}. Note that the PDF
does not need to be normalised, \ie $F_{n - 1}(x_n)$ is not required
to be unity, because the \cmd{CDF} class returns $F_i(x)/F_{n -
  1}(x_n)$.

The \cmd{InverseCDF} class is also initialised with a \cmd{PDF}
object, but returns the inverse CDF of the PDF as follows.
\begin{enumerate}
\item The $y$-values of the CDF are built by integrating the
  interpolating function over each interval from the PDF using
  \equ{cdf}.
\item If limits are passed, \eg the $x$-limits over which the inverse
  CDF is defined, the CDF values for these points are stored as $u_0$
  and $u_1$ where $0 \leq u_0,u_1 \leq 1$.
\item The $x$ and $y$-values are swapped.
\item When called with an argument $u$ between $0$ and $1$, the
  argument is transformed as $u' = u_0 + u(u_1 - u_0)$ to fall within
  the $x$-limits.
\item The modified regula falsi bin finding of the base
  \cmd{Histogram} class is used to determine interpolating interval
  $i$ for $u'$.
\item Root finding using \equ{quartic} for interval $i$ is performed,
  given $u'$, and the resulting $x$ is returned along with the weight
  $u_1 - u_0$. If no $x$-limits have been required, this weight is
  unity.
\item If the analytic solution fails numerically, \ie $x < x_i$ or $x
  > x_{i + 1}$ the companion matrix method is used, as implemented in
  the \cmd{roots} method of the \cmd{numpy} package. If \cmd{numpy} is
  not available, linear interpolation is used.
\end{enumerate}
The first three steps are only performed at initialisation of
\cmd{InverseCDF}, while the remaining steps are performed each time an
inverse CDF value is calculated.

The \cmd{Sample1D} class returns an $x$-value sampled from an
interpolated PDF, given a uniform random number between $0$ and
$1$. This is done by constructing an internal \cmd{InverseCDF} object
and using this to calculate $x$ given $u$. Optionally, $x$-limits on
the sampling range can be passed. The \cmd{Sample2D} class is similar
to \cmd{Sample1D} but now returns a sampled $x$ and $y$-value with an
associated weight using the method of \scn{bs} as follows.
\begin{enumerate}
\item A \cmd{InverseCDF} object is constructed for the projected PDF,
  given a \cmd{PDF} object for each $y$-value. The interpolating
  coefficients are calculated with \equ{projection}.
\item When passed two uniform random numbers, $u$ and $v$, an
  $x$-value and its corresponding weight $w_x$ is sampled from the
  projected \cmd{InverseCDF}.
\item Given this $x$, a \cmd{PDF} and subsequent \cmd{InverseCDF}, \ie $G^{-1}(v|x)$, is built and a $y$-value with an associated weight $w_y$ is sampled.
\item The $x$ and $y$-values are returned with a weight of $w_xw_y$.
\end{enumerate}
If no $y$-limit is placed on the sampling, the returned weight will be
constant, and if additionally no $x$-limit is required, the returned
weight will be unity. The first step is performed only during
initialisation, while the remaining steps are performed for each
sampling. The \cmd{SampleTH1} and \cmd{SampleTH2} classes provide
convenient wrappers for sampling one and two-dimensional histograms
from the \cmd{ROOT} package~\cite{Brun:1997pa}.

The \cmd{PythiaDecay} class takes three-momentum vectors sampled with
the \cmd{ParticleGun} class and performs particle decays using the
\cmd{pythia8} module. The \cmd{ParticleGun} class uses the
\cmd{Sample2D} class and the interpolation grids described in
\scn{pgun} to generate momentum three-vectors for a given particle
type and set of sampling grids.
\begin{enumerate}
\item A PDF bivariate in $\rho$ and $\eta$ is built from the requested
  histogram and used to create an \cmd{InverseCDF} object.
\item Because the PDF at $\rho = 0$ must be zero, \ie particles are
  not produced with infinite transverse momentum, an additional point
  with a value of zero is inserted for each $\eta$ value with $\rho =
  0$.
\item Similarly, an additional point with a linearly extrapolated
  value is inserted for each $\eta$ value with $\rho = 1$, if such a
  point does not already exist. This value is required to be $\geq 0$.
\item To generate a momentum three-vector, a value for $\rho$ and
  $\eta$ is sampled from the \cmd{InverseCDF}.
\item The transverse momentum is calculated as $\rho^{-1/k} - 1~\gev +
  \pt_{\min}$.
\item The azimuthal angle $\phi$ is uniformly sampled between $0$ and
  $2\pi$, while $\cos(\theta) = (e^{2\eta} - 1)(e^{2\eta} - 1)$ and
  $\sin(\theta) = \sqrt{1-\cos^2(\theta)}$.
\item The three-momentum is returned as $[\pt\cos(\phi),
  \pt\sin(\phi), \pt\cos(\theta)/\sin(\theta)]$ with a weight of
  $w_\eta$. The $\rho$-weight is included in the cross-section value.
\end{enumerate}
The member \cmd{sigma} of the \cmd{ParticleGun} class provides the
cross-section for the sampling in millibarn, given the \pt-limits
passed by the user, \ie accounting for the $\rho$-weight. This is to
ensure that sampling is not performed when the \pt-limits produce a
null interpolated phase-space. The same can not be done for the
$\eta$-limits as the $\eta$-weight may be variable, given
$\rho$. However, if the projected PDF in $\rho$ is non-zero, then for
the sampled $\rho$ the $\eta$ PDF is guaranteed to have non-zero bins.

\section{Validation, Performance, and Applications}\label{sec:vpa}

\begin{figure}
  \includegraphics[width=0.5\textwidth]{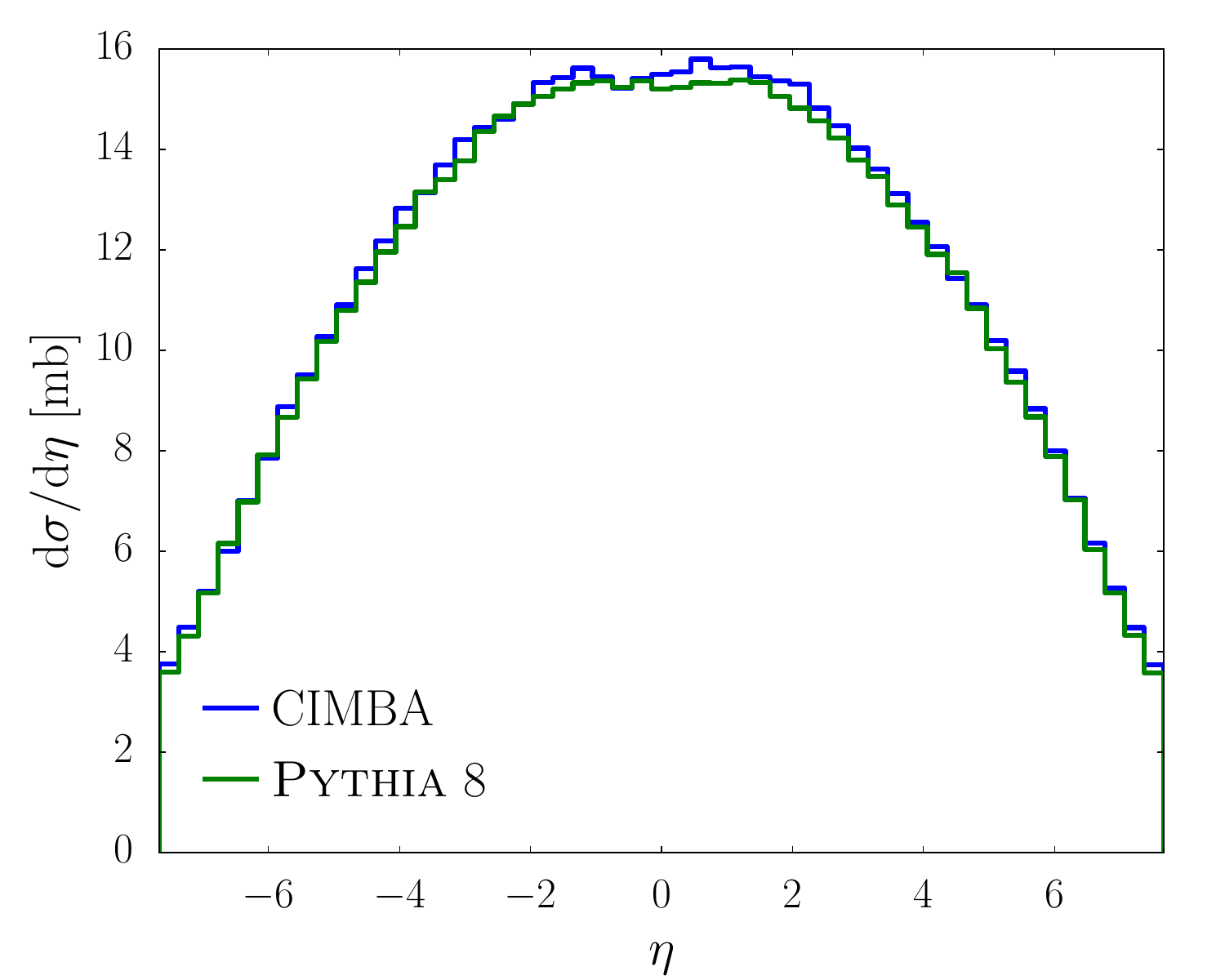}
  \includegraphics[width=0.5\textwidth]{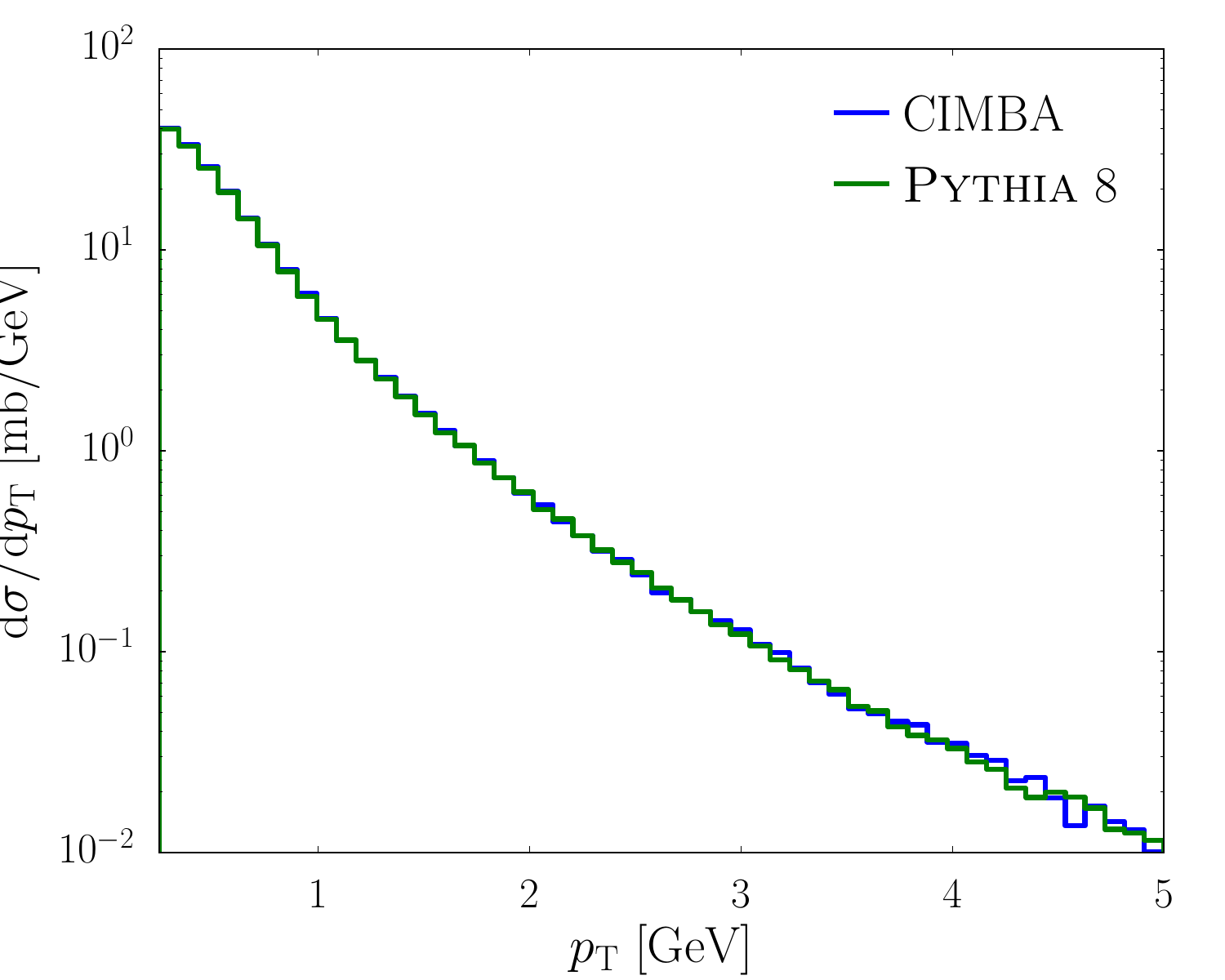}
  \caption{Comparison of \cimba and \pythia generated (left)
    pseudorapidity and (right) transverse momentum distributions for
    $\pi^+$ production at the LHC.\label{fig:validate}}
\end{figure}

The pseudorapidity and transverse momentum distributions of positively
charged pions, $\pi^+$, produced by \cimba and \pythia are compared in
\fig{validate}. There is good agreement not only in the shape of the
distributions, but also in the absolute normalisations, \ie the
predicted differential cross-sections. Tests have been performed for
other particle species with similar levels of agreement
observed. \cimba can generate roughly 25 particles in the time that
\pythia produces a single soft QCD event. In a typical LHC event,
$\approx 30$ $\pi^+$ are produced on average with no fiducial
requirements, and so for this worst case scenario, generation time for
$30$ $\pi^+$ is roughly equivalent between \cimba and
\pythia. Realistic fiducial regions rapidly reduce the event
multiplicity, \eg requiring $\pt > 1~\gev$ already reduces the $\pi^+$
multiplicity to $\approx 2$.

Six grid files, each generated from $10^9$ \pythia events, are
provided with \cimba.
\begin{itemize}
  \item \texttt{pp7TeV}: soft QCD at the LHC with $\sqrt{s} = 7~\tev$.
  \item \texttt{pp8TeV}: soft QCD at the LHC with $\sqrt{s} = 8~\tev$.
  \item \texttt{pp13TeV}: soft QCD at the LHC with $\sqrt{s} =
    13~\tev$.
  \item \texttt{pp14TeV}: soft QCD at the LHC with $\sqrt{s} =
    14~\tev$.
  \item \texttt{pp27GeV}: soft QCD at SHiP with a proton beam energy
    of $400~\gev$.
  \item \texttt{ppbar980GeV}: soft QCD at the Tevatron with $\sqrt{s}
    = 980~\gev$.
\end{itemize}
For each grid file, two sets of grids are provided: \texttt{all} which
includes the final version of all particles produced in the event, and
\texttt{had} where only particles produced directly from hadronisation
are kept. From these grids, all single particle observables, for any
particle species produced by \pythia, can be generated. Observables
requiring particle correlations cannot be produced, \eg jets. As
detailed in \scn{pgun}, code is provided so new grids with different
configurations may be generated by the user. Additionally, commonly
used grids will be added to \cimba upon request.

In the following examples, the estimated performance of \cimba does
not include the cost of generating the necessary interpolation
grids. In principle, the entirety of the \pythia samples used to
generate the \cimba grids could be stored to disk. However, using
naive storage techniques, this would require over $60~\mathrm{TB}$ of
disk space per grid file. Simply accessing and storing such data would
be computationally prohibitive. Reduced data could also be stored, \ie
four-vectors for a specific particle species, but then \pythia
generation would be necessary whenever a new particle species was
required. Finally, some particle species are produced so rarely by
\pythia that smoothing would still be required when generating
observables; \cimba handles this transparently.

\begin{figure}
  \includegraphics[width=0.5\textwidth]{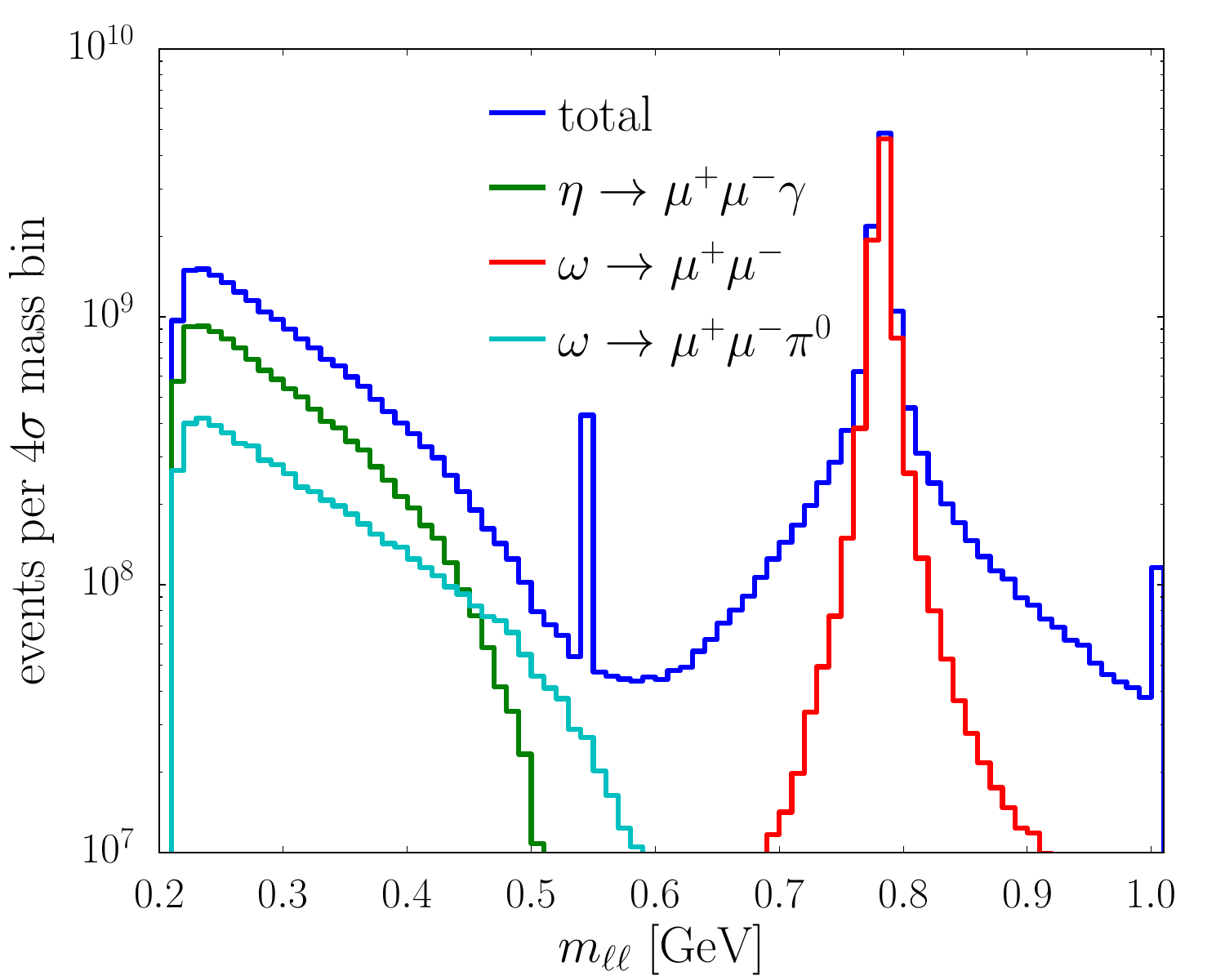}
  \includegraphics[width=0.5\textwidth]{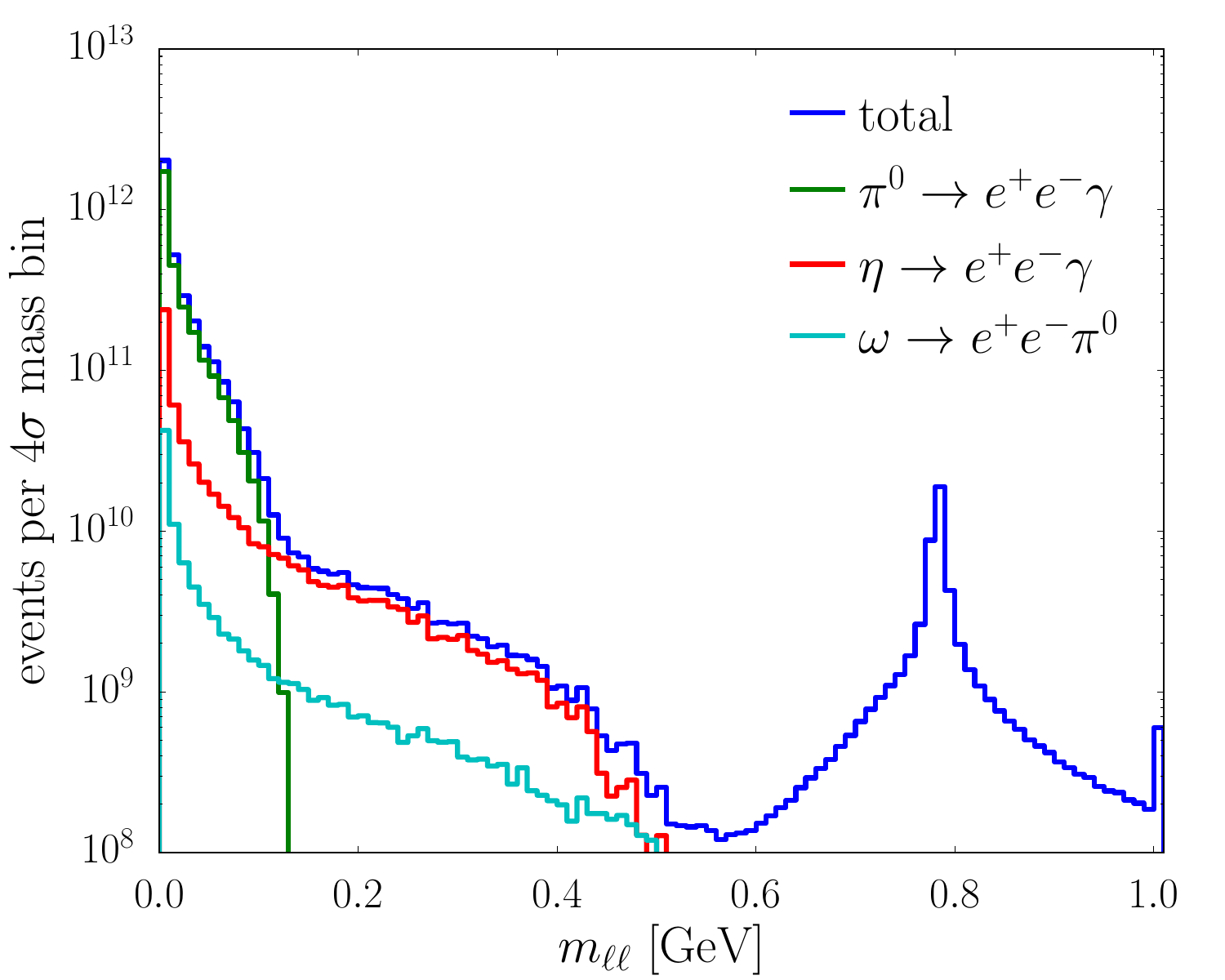}
  \caption{\cimba generated (right) di-muon mass spectrum for LHCb Run
    3 following the proposed dark photon search of \rfr{Ilten:2016tkc}
    and (left) di-electron mass spectrum following the proposed true
    muonium search of \rfr{CidVidal:2019qub}.\label{fig:example}}
\end{figure}

Two examples are given in \fig{example}, demonstrating use cases for
\cimba. In the inclusive di-muon dark photon search proposal using
LHCb Run 3 data of \rfr{Ilten:2016tkc}, the dark-photon production
rate is proportional to the expected electromagnetic (EM) background
which is produced primarily from $\eta$, $\omega$, and $\rho$-meson
decays, for dark photon masses less than $1~\gev$. The results of
\cimba in \fig{example} agree well with the results of
\rfr{Ilten:2016tkc}.\footnote{The distributions of \fig{example} and
  Figure~2 from \rfr{Ilten:2016tkc} do not match exactly. The $\eta$
  and $\rho$ contributions entering the distribution of
  \rfr{Ilten:2016tkc} were scaled to better match LHC
  data. Additionally, different versions of \pythia have been used,
  and a smoothing function was applied to the distribution of
  \rfr{Ilten:2016tkc}.} Within the fiducial region of this search, on
average $\approx 0.1$ $\eta$ mesons are produced per event and so
using \cimba corresponds to an $\approx 250\times$ speedup. A sample
which would previously take on the order of a week to produce, can now
be generated in roughly a half hour. This is particularly useful for
determining expected experimental reach for new physics produced from
light mesons. Similarly, when recasting dark photon results to
non-minimal models~\cite{Ilten:2018crw} the ratio between production
mechanisms is needed, which can now be quickly calculated with \cimba.

The displaced background for the inclusive di-muon dark photon search
of \rfr{Ilten:2016tkc} at low dark photon masses is primarily from
semi-leptonic $B$-meson decays of the form $B \to \mu D [\to \mu
  X]$. Only $< 0.01$ $B$-mesons per event are in the fiducial region
of the search; generation with \cimba results in a $> 2500\times$
speedup. Of course, a more strategic generation of events could be
performed with \pythia, \eg generating \cmd{HardQCD:hardbbbar} events
with minimum transverse momentum cuts. However, strategies such as
this oftentimes result in the loss of some physics processes, in this
case the production of $B$-mesons from gluon splitting in events
without a hard $b\bar{b}$ process. In the LHCb dark photon search of
\rfr{Aaij:2017rft}, an implementation of \rfr{Ilten:2016tkc}, large
generated samples of $B \to \mu D [\to \mu X]$ events were required to
create a displaced background template. With \cimba, higher statistics
and more detailed generation of this background can be achieved.

The $^3S_1$ state of true muonium, a bound di-muon atom, can be
produced in a similar fashion to dark photons, with the notable
exception that true muonium has a small binding energy of $1.4~\kev$
and will typically dissociate upon impact with detector material. In
\rfr{CidVidal:2019qub} a search proposal for true muonium using
di-electron final states from $\eta$-meson decays in LHCb Run 3 data
is outlined. In \fig{example} the di-electron mass distribution is
given for this search proposal, as generated by \cimba. Note that
there is no $\eta \to e e$ peak, unlike the $\eta \to \mu \mu$ peak,
as this decay channel is helicity suppressed by a factor of
$m_e^2/m_\mu^2$. The result agrees well with the prediction of
$\approx 3400$ TM events from \rfr{CidVidal:2019qub}, given a signal
to EM background ratio of $1.2\times10^{-6}$ for a $4\sigma$ mass
bin.\footnote{A similar distribution to the di-electron distribution
  of \fig{example} is not available in \rfr{CidVidal:2019qub} as
  generating this distribution was too time consuming using
  conventional methods. Additionally, for \rfr{CidVidal:2019qub}, only
  a single mass point is required, \ie at the mass of true muonium.}
Here, the full mass distribution is given, which can also be used to
determine a naive reach for dark photon searches with di-electron
final states.

Both the dark photon and true muonium examples are provided in the
\cmd{dilepton.py} example in the \cimba package. After particle
generation with \cimba, decays are performed using \pythia. All
relevant di-lepton decays are determined using the \pythia decay
tables. In this example \cimba automatically estimates the branching
fractions for missing di-lepton channels in \pythia by scaling the
branching ratio for the equivalent channel with a photon by a factor
of $\alpha$, neglecting phase-space suppression. For example, the
decay $\eta \to \mu^+ \mu^- \pi^+ \pi^-$ is missing in \pythia, so its
branching fraction is calculated from the channel $\eta \to \gamma
\pi^+ \pi^-$. In this particular case, the branching ratio is
estimated to be $3.4\times10^{-4}$ but the expected branching ratio is
$1.2\times10^{-8}$~\cite{Faessler:1999de} and so this channel is added
explicitly.

No explicit attempt has been made at evaluating the uncertainty
introduced via the interpolation methods of \cimba. This is because
\cimba is intended for use within the non-perturbative QCD regime,
where the tune of the event generator producing the interpolation
grids is expected to be the primary source of systematic
uncertainty. However, the uncertainty introduced via interpolation can
be estimated by re-binning the interpolation grids, \eg halve the
number of bins, and calculating the difference in shapes between the
fine and coarse interpolation grids.

\section{Conclusions}\label{sec:con}

Generating kinematics for single particles produced in simulated
minimum bias events at the LHC within well defined fiducial regions
can be computationally expensive, particularly for particle types
rarely produced in the hadronisation process. This can be problematic
for experiments such as LHCb, where the majority of simulation samples
are extracted from generated minimum bias events. The \cimba package
provides a lightweight and fast particle gun method to generate smooth
interpolated distributions of such particles with well defined
fiducial regions in pseudorapidity and transverse momentum. The
efficacy of \cimba is demonstrated by reproducing the results from
proposed dark photon and true muonium searches with LHCb. These
results were generated on the order of minutes, rather than the
previous computation time on the order of weeks. Additionally, this
method for sampling distributions from cubic interpolated histograms,
both univariate and bivariate, has applications outside of particle
physics where smooth sampling is required from non-parametric
distributions.

\section{Acknowledgements}

We thank Stephen Farry, Jonathan Plews, Yotam Soreq, and Nigel Watson
for providing useful feedback and testing. PI is supported by a
Birmingham Fellowship.

\section*{Bibliography}
\bibliographystyle{elsarticle-num}
\bibliography{cimba}
\end{document}